# WiP: A Novel Blockchain-based Trust Model for Cloud Identity Management

Keltoum Bendiab
*Department of Electronics,
Sciences of Technology faculty
frères Mentouri university
Constantine, Algeria*
bendiab.keltoum@umc.edu.dz

Nicholas Kolokotronis
*Department of Informatics and
Telecommunications
University of Peloponnese
Tripolis, Greece*
nkolok@uop.gr

Stavros Shiaeles
*Centre for Security, Communications
and Network Research (CSCAN)
University of Plymouth
Plymouth, UK*
stavros.shiaeles@plymouth.ac.uk

Samia Boucherkha
*LIRE laboratory
Abdelhamid Mehri University
Constantine, Algeria*
samchicki@yahoo.fr

*Abstract*—Secure and reliable management of identities has become one of the greatest challenges facing cloud computing today, mainly due to the huge number of new cloud-based applications generated by this model, which means more user accounts, passwords, and personal information to provision, monitor, and secure. Currently, identity federation is the most useful solution to overcome the aforementioned issues and simplify the user experience by allowing efficient authentication mechanisms and use of identity information from data distributed across multiple domains. However, this approach creates considerable complexity in managing trust relationships for both the cloud service providers and their clients. Poor management of trust in federated identity management systems brings with it many security, privacy and interoperability issues, which contributes to the reluctance of organizations to move their critical identity data to the cloud. In this paper, we aim to address these issues by introducing a novel trust and identity management model based on the Blockchain for cloud identity management with security and privacy improvements

## I. INTRODUCTION

Since cloud computing gains more and more importance, Identity management in and across clouds has become an area of great concern in this domain [1], [2]. According to Gartner [3], identity management is amongst the top three challenges that organizations face while moving their applications to a cloud-based model. Federated Identity Management (FIM) is the most used solution to overcome these issues; especially the cross-domain issues [2], [4], [5]. This approach simplifies the user experience by using cross-domain Single Sign-On features [5], and reduces the costs of administrating user accounts [4]. The federated exchange of user identity information between the separate security domains is based on standardized protocols, such as Security Assertion Markup Language, Shibboleth, WS-Federation, Liberty Alliance, OpenID and OAuth. All these frameworks usually follow a similar architectural concept [6], basically involving Identity Providers (IdP) and Service Providers (SP) in a structure called Circle of Trust, where IdPs and SPs have to trust each other; in particular, IdPs have to trust the SPs to securely handle a user's identity data [3], whereas the SPs have to trust the IdPs to correctly authenticate users that want to access its services and protected resources [5]. While identity federation seems to be a promising approach for adopting identity management in cloud computing, the underlining trust model is poorly defined [3], [7], [8] and manually managed by pre-configured Trust Anchor Lists with a Public Key Infrastructure [9]. In this model, trust is established based on business agreements that must be set well before the interactions take place, which leads in forming closed and isolated communities [10]. Such a trust model is unsuitable for dynamic environments such as cloud computing, in which trust between parties involved in a federation process should be created dynamically on demand instead of being statically defined and manually managed by a central authority. Researches generally believe that decentralised and dynamic management of trust is vital for ensuring safe, secure and transparent use of user's sensitive data and cloud services [2], [10]. In addition, identity federation carries significant security and tracking risks [11], and does not have a good approach to protect user's privacy [12]. The sharing of users' identity attributes among different entities involves the collection of user data without user consent [2], [12]. Many studies show that the personal data collected can be misused by malicious SPs and IdPs [2], or compromised [12], or otherwise improperly disclosed [11], [12], which could lead to a higher level of information leakage [2], [12], [11]. These issues will become even more complicated with cloud computing [2] since it is a highly dynamic, multi-tenancy, insecure, and open environment. The aforementioned drawbacks of FIM constitute the main obstacle to the successful adoption of this approach in cloud computing and emphasize the need for a new cloud identity management scheme with improved security and privacy. Motivated by the above, we aim to address these challenges by proposing in this paper a distributed trust and identity model that relies on the Blockchain. The proposed trust model allows the cloud service providers (CSPs) to autonomously manage their trust relationships in a dynamic and distributed manner without the need for centralized authorities, such as the IdPs. This makes cloud service provisioning and user interaction easier, more secure, and more flexible.

The rest of the paper is organized as follows: Section II presents basic concepts related to Blockchains for identity management. Section III discusses some related works and their limitations. In section IV we introduce our trust model. Section V introduces an application example. Finally, section VI reviews the content of the paper, presents the conclusions and outlines our future work.

*This work was supported by CYBER-TRUST project, which has received funding from the European Union's Horizon 2020 research and innovation programme under grant agreement no. 786698.*

## II. BLOCKCHAIN FOR CLOUD IDENTITY MANAGEMENT

In recent years, blockchain technology has attracted a lot of attention from Internet users, researchers and stakeholders across a wide range of sectors like finance [13], healthcare [14], logistics and transportation [15], defense and the government sector [16]. First used as a public ledger for the Bitcoin cryptocurrency [17], blockchain is regarded as a distributed, fault-tolerant and trusted data structure that is replicated and shared among the members of a peer-to-peer network [18]. The database is made of a linear sequence of chained blocks; each one represents a single data unit, typically made of a header of metadata, a record of transactions and the cryptographic hashes corresponding to the previous and current block to ensure continuity and immutability [18]. New blocks are added at the end of the chain. A simplified representation of a blockchain architecture is given in Figure 1.

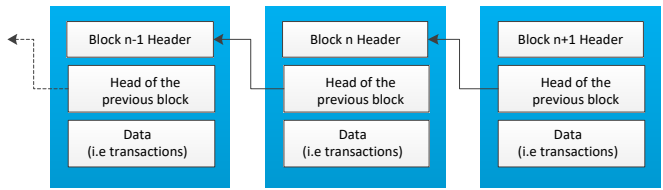

Fig. 1. Blockchain architecture.

The transparent, distributed and trusted nature of the blockchain enables the development of a non-refutable and unforgeable ledger of data [19], which is the key feature of a many successful applications, such as identity management. Combining this new technology of trust with cloud identity management can offer considerable advantages compared to FIM approach, and solve some, if not all, problems mentioned above. The works carried out in [19], [20] and [21] show that the blockchain provides a promising opportunity to standardize and strengthen identity management processes. It can potentially assist in improving transparency and security in cloud identity management [22]. In fact, the use of public blockchains renders it difficult to compromise the integrity of their records without being noticed by the entire network [23]. In addition, the use of distributed (and in many cases fault-tolerant) consensus protocols allows preventing malicious activities, like double-spending, hacking [23], and mitigating identity theft and fraud [22]. Moreover, removing intermediaries (e.g. central authorities), by using the blockchain, has the potential to reduce the costs, time, and complexity of cloud identity management [19], [20]. All these characteristics demonstrate that blockchain technology seems to be the perfect match for cloud identity management needs.

## III. RELATED WORKS

In last few years, numerous works have emerged for adapting blockchain technology in order to meet new identity management requirements. In this context, the authors in [21] proposed a decentralized personal data management platform by combining blockchain technology with an off-blockchain storage solution. The proposed framework focuses on user's privacy issues on mobile platforms (social networks and big data), where service providers collect users' sensitive data without their knowledge or control. This work shows the advantages of blockchain technology in enhancing user privacy in the context of social networks and big data.

Towards the same direction, a new blockchain-based authentication and trust management model, called BATM, was proposed in [24]. The BATM blockchain is used as a distributed database to store authentication public keys, digital signatures and information about peers. The proposed framework helps to ensure validity and integrity of cryptographic authentication data and associate a peer trust level in the area of decentralized ad-hoc networks. Another application of blockchain in identity management was proposed in [25], where a new public and decentralized authentication scheme, named *Certcoin*, was presented. The Certcoin blockchain is built on top of Namecoin [26] which is a cryptocurrency designed to act as a secure decentralized namespace (DNS). Certcoin provides a decentralized public key infrastructure in order to maintain a public ledger of the domains and their associated public keys. This solution helps to mitigate many of the issues in current PKIs, such as the need for a trusted third party and the lack of efficient key retrieval mechanisms. As trust is a vital component of identity management, *TrustChain* [27] is another use of the blockchain technology in this area, with a focus on building trust among users (individuals or organizations) with no prior trust relationship. It is a distributed trust management solution that helps in determining trustworthiness of agents in an online community. TrustChain provides a new scalable and Sybil attack resistant blockchain solution.

These schemes presented above demonstrate the use of blockchain for identity management and its potential to enhance security and privacy compared to traditional approaches. However, each scheme is specifically designed for a specific area. To our knowledge, the use of a blockchain for identity management in cloud environment has never been fully explored. In this work, we aim to integrate blockchain in the cloud identity management by proposing a simple and fully decentralized blockchain-based trust and identity management solution. The details of the proposed system are provided in the following sections.

## IV. PROPOSED SOLUTION

### A. System architecture

First, we begin with an overview of the architecture of our system. As shown in Figure 2, the three critical entities involved in the proposed solution are cloud users, Cloud Service Providers (CSP) and the Trust Management Platform (TMP), whose roles are explained below.

*1) Cloud user:* is the entity that wants to access services and protected resources from different CSPs in the TMP. The user is registered in one or more CSPs in the TMP, called home CSP. Unregistered users in the TMP will have to go through a registration phase to obtain their digital identities.

*2) Cloud service provider:* The CSP offers various cloud services and resources to cloud users and requires proper identification and authentication. The user's home CSP is the

CSP responsible for authenticating the user. It is assumed that this CSP is ultimately trusted by the user.

3) *Trust management platform:* The TMP incorporates the Blockchain network, which consists of a number of participating CSPs. All authentication transaction records will be recorded in the form of blocks and be verified by the blockchain nodes (i.e. the CSPs). The TMP is dynamically extensible and a new CSP can join it automatically in runtime. That allows rapid and seamless interactions between multiple unknown CSPs without the need for preconfigured trust relationships. This leads to increase significantly the scalability and flexibility needed for the cloud identity management.

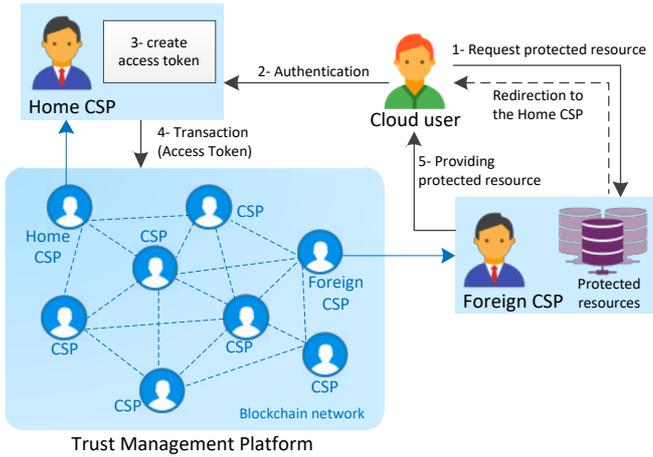

Fig. 2. Architectural overview.

The core idea of our system is illustrated in Figure 2. It mainly involves the following steps: First, the user wants to access protected resources and services of a CSP in the TMP with which he is not registered with (step 1); such CSPs will be called foreign CSPs. In this case, the foreign CSP redirects the user to the home CSP in order to perform authentication. Then (steps 2, 3), the home CSP validates the authentication request and creates an access token that contains claims about the identity and privileges associated with the user's account.

The token is stored in the blockchain, in order to prove that this is a valid token owned by the user and issued by a trusted CSP in the TMP, thus significantly increasing the security and level of the exchange. Once the token is validated and stored in the blockchain (step 4), the CSP will continue with the transaction and allow the user to access protected resources based on the specified access privileges (step 5); otherwise, the CSP will stop the transaction and deny access to the protected resources.

### B. Blockchain architecture

As mentioned before, in the proposed framework, the blockchain is considered as a decentralized trust model that allows CSPs managing their trust relationships in a distributed manner without relying on a trusted third-party (i.e. the IdPs), while it is also used as a public shared ledger that stores all the transactions among the interacting CSPs. The transactions are used to create and store the tokens that authenticate users to foreign CSPs and guarantee their access to protected resources. An access token TKN contains claims about the identity and privileges associated with a user account $u$, defined by the creator $H_{csp}$ of the token (i.e. the home CSP) and sent to the recipient $F_{csp}$ (the foreign CSP); the protected resources $res$, for which access is requested by the user, are identified by their address $Addr(res)$. Each token corresponds at most to one transaction TX, whose structure is represented as follows

$$TX = (TXID \,||\, N_{in} \,||\, V_{in}[] \,||\, N_{out} \,||\, V_{out}[])$$

where TXID is the identifier of the transaction, $N_{in}$ and $N_{out}$ hold the number of transaction inputs and outputs, whereas $V_{in}$ and $V_{out}$ are the input and output vectors. Each input in $V_{in}$ has the following components: the index $idx$ of the input in $V_{in}$, the reference to previous input $ref_{in}$, the encrypted user information $Enc(U)$, and the address $Addr(res)$ of the requested resource. Likewise, each output in $V_{out}$ has the following components: the index $idx$ of the output in $V_{out}$, the reference to the previous output $ref_{out}$, the created token *TKN*, and the address $Addr(F_{csp})$ of the recipient CSP ($F_{csp}$). To store the tokens in a chronological order, transactions are linked to each other by including a hash of the previous transaction into a field of the current transaction. Identification of the CSPs in the TMP and the authenticity of the tokens are guaranteed by digital keys, signatures and addresses. Each CSP is associated with a pair keys ($prv_{csp}$, $pub_{csp}$), where $prv_{csp}$ denotes the private key and $pub_{csp}$ denotes the public key. The private key $prv_{csp}$ is used to create signature called $sig$ that is required to sign the transaction and prove the authenticity of the token. The $H_{csp}$ encrypts the user information, which limits access to the user's identity only to the key holder. Keys are derived in a tree structure, so that a parent key can derive a sequence of children keys and implemented by using the Elliptic Curve Digital Signature Algorithm (ECDSA) with secp256k1 curve [28].

### C. Consensus protocol

The validation of transactions is performed by the CSPs in the network, who are responsible for maintaining the ledger's integrity as well. The transactions (carrying the tokens) are not immediately added to the blockchain; instead, they are packed into blocks containing many transactions for efficiency reasons. This allows to avoid flooding the network with each generated block, and to minimize the time spent on block generation. The algorithms that are typically used by the peers to validate the authenticity of transactions are called consensus protocols. There have been many proposals in this area (see e.g. [18] and the references therein for an overview), but our approach relies a Proof of Stake (PoS) solution.

In the proposed solution, each new transaction is broadcast to all network peers (i.e. the CSPs) and a new block is generated from the CSP that is eligible to do so. The eligibility and other aspects of the PoS protocol are defined by the following high-level functions:

*1)* CheckEligibility($h_{blk}$, $d$, $key_{csp}$, $stake_{csp}$). The function is used to elect a leader for generating the next block in the chain by checking if a CSP is eligible. Its input arguments are the last block header $h_{blk}$, a target value $d$, the account key $key_{csp}$, (different from the public/private key pair used in other operations) and the stake or the assets $stake_{csp}$ owned by the CSP; it returns a Boolean value (true or false).

*2) GenerateBlock(blk, d).* The function is used for generating the block to store on the blockchain. Its input arguments are the transactions block $blk$ and a target value $d$; it returns a pair $\langle prf, sig \rangle$ of values, where $prf$ is a proof-of-eligibility (that other CSPs can verify) and $sig$ is the digital signature of the block. These values are obtained as follows

$$prf = \text{Hash}(pub_{csp} \| prf_{old})$$
$$sig = \text{Sign}(prv_{csp}, h_{blk})$$

where $pub_{csp}$ and $prv_{csp}$ are the public and private key of the CSP generating the new block, while $prf_{old}$ is the last proof-of-eligibility (like Nxt's PoS protocol). The function returns empty when CheckEligibility indicates that the CSP is not eligible.

*3) ValidateBlock(blk, d, prf).* The function is used for verifying a block having been stored on the blockchain. It returns true only if CheckEligibility is true and $blk$ (the signature of the block), $prf$ (the proof) are both valid.

*4) Resolve(fork₁, ..., forkₙ).* The function is utilized for resolving possible cases where multiple forks are present by returning a unique fork, say $fork_l$, to work with. This situation occurs when many CSPs are simultaneously eligible for generating the next block.

The system-wide target value $d$ that has been used in the above functions controls the block generation time (by adjusting the difficulty of the problem that a CSP has to solve), which in our application scenario is envisaged to be 30 seconds. The difficulty of the problem to be solved by each CSP also depends on the time $time_{csp}$ elapsed since the last block generated by the particular CSP, its stake $stake_{csp}$, as well as its trust value $t_{csp}$ in the TMP. The latter also gives incentives to the CSPs to adhere to the specified protocol; the expression

$$d_{csp} = d \cdot time_{csp} \cdot stake_{csp} \cdot t_{csp}$$

provides a simple way to define a CSP-specific difficulty level for the PoS protocol designed for cloud identity management. Then, the eligibility of a CSP for being a leader is decided by checking whether the condition Prefix($prf$, $k$) < $d_{csp}$ holds for a system-wide parameter $k$ (e.g. 64 or 128).

### D. Trust model

In order to ensure effective and more secure interactions among the participating CSPs in the TMP, transactions stored in the blockchain are used to evaluate the trustworthiness of each CSP over time based on their behavior. In the TMP, each CSP has two kinds of roles: home CSP ($F_{csp}$) and foreign CSP ($F_{csp}$). A home CSP provides authentication services of registered users ($u$) to another CSP in the TMP, while, a foreign CSP grants access to protected resources to registered users with another CSP in the TMP. In order to calculate the trustworthiness of each CSP, we define $Trust_{n,t}(CSP)$ to be the trust level of the CSP up to $n$ transactions in the time interval $t$. It is in the range $0 \leq Trust \leq 1$ and depends on the features that are presented in table 1.

*1) User credibility:* we use $Cred_{n,t}(u)$ to denote the credibility value that CSPs in the TMP give to the user up to n transactions in the time interval t. It is defined as follows

TABLE I. FEATURES OF TRUSTWORTHINESS

| Trust feature | Description |
|---|---|
| Credibility (Cred) | It refers to the users credibility according to the behavior exhibited in transactions made with other CSPs in the TMP. This feature helps to identify malicious users and users' ability in delivering truthful feedback about the quality of the CSPs that provided services. Feedbacks provided by users with higher credibility are more trustful than those from users with lower credibility. |
| Authentication (Auth) | It refers to the ability of $H_{csp}$ in fulfilling authentication requirements by registers and authenticates users safely and securely. It is derived from the past behavior of the $H_{csp}$ registered users reported by other CSPs in the TMP. |
| Satisfaction (SAT) | It refers to the degree of the satisfaction that the CSPs in the TMP have about a given $F_{csp}$ based on its services quality. It keeps a record of the satisfaction degree for all the transactions that a $F_{csp}$ makes with the other CSPs. |

$$Cred_{n,t}(u) = \frac{1}{k} \cdot \sum_{i=1}^{k} Cred_{n,t}(F_{csp}(i), u) \quad (1).$$

Where, $Cred_{n,t}(F_{csp}, u)$ represents the credibility value that $F_{csp}$ gives to the user up to $n$ transactions in the time interval $t$; the initial value is set to one ($Cred_{0,0}(F_{csp}, u) = 1$) and the credibility update function is defined in equation (2).

$$Cred_{n,t}(F_{csp}, u) = \frac{Trust_{n-1,t}(F_{csp}) \times Cred_{curr} + Cred_{n-1,t}(F_{csp}, u)}{2} \quad (2).$$

Where, $Trust_{n-1,t}(F_{csp})$ represents the trust level of the $F_{csp}$ up to n-1 transactions in the time interval t, and $Cred_{curr}$ represents the user $u$ current credibility value given by the $F_{csp}$. It is computed on the basis of the feedback provided by the $F_{csp}$ to the user in the transaction based its behavior. It is defined as

$$Cred_{curr}(u) = \begin{cases} 0 \leq x \leq 0.20 & \text{If very bad} \\ 0.20 < x \leq 0.40 & \text{If bad} \\ 0.40 < x \leq 0.60 & \text{If medium} \\ 0.60 < x \leq 0.80 & \text{If good} \\ 0.80 < x \leq 1 & \text{If excellent} \end{cases}$$

*2) Authentication:* we use $Auth_{n,t}(H_{csp})$ to denote the authentication level of the $H_{csp}$ given by the other CSPs based on its authentication services up to $n$ transactions in the time interval $t$. It keeps a record of the authentication level of all the transactions a $H_{csp}$ makes with other CSPs participating in the TMP. It is computed as follows

$$Auth_{n,t}(H_{csp}) = \frac{1}{k}\sum_{i=1}^{i=k} Auth(F_{csp}(i), H_{csp}) \quad (3).$$

Where $Auth_{n,t}(F_{csp}, H_{csp})$ represents the authentication level the $F_{csp}$ gives to the $H_{csp}$ based on its authentication service up to $n$ transactions in the time interval $t$. Its initial value is equal to $Auth_{0,0}(F_{csp}, H_{csp}) = 0$. The authentication update function is defined as follows

$$Auth_{n,t}(F_{csp}, H_{csp}) = \frac{Auth_{curr} + Athen_{n-1,t}(F_{csp}, H_{csp})}{2} \quad (4).$$

Where, $Auth_{curr}$ represents the authentication level for the current transaction. It is derived from the feedback provided by the $F_{CSP}$ to the user of the $H_{csp}$ in the transaction based on its behavior. It is defined as follows

$$Auth_{curr}(H_{csp}) = Cred_{curr}(u) \quad (5).$$

3) *Satisfaction:* we use $SAT_{n,t}(F_{csp})$ to denote the degree of satisfaction that other CSPs have upon the $F_{csp}$ based on its quality of services up to $n$ transactions in the time interval $t$. It is computed by applying the calculation rule in equation (4).

$$SAT_{n,t}(F_{csp}) = \frac{1}{k}\sum_{i=1}^{i=k} SAT_{n,t}(H_{csp}(i), F_{csp}) \quad (4).$$

Where, $SAT_{n,t}(H_{csp}, F_{csp})$ represents the degree of satisfaction that $H_{csp}$ has upon the $F_{csp}$ based on its quality of service up to $n$ transactions in the time interval $t$, and $k$ is the number of CSPs in the TMP; its initial value is $SAT_{0,0}(H_{csp}, F_{csp}) = 0$. The satisfaction update function is defined in equation (5)..

$$SAT_{n,t}(H_{csp}, F_{csp}) = Cred_{n,t}(u) \times SAT_{curr} + (1 - Cred_{n,t}(u)) \times SAT_{n-1,t}(H_{csp}, F_{csp}) \quad (5).$$

Where, $SAT_{curr}$ is the satisfaction value for the current transaction. The value of $SAT_{curr}$ is computed based on a feedback system, where the user $u$ from the $H_{csp}$ rates the quality of service offered by $F_{csp}$ after each transaction according to the following rule

$$SAT_{curr}(F_{csp}) = \begin{cases} 0.00 \leq x \leq 0.2 & \text{If fully dissatisfied} \\ 0.20 < x \leq 0.45 & \text{If dissatisfied} \\ 0.45 < x \leq 0.6 & \text{If partially satisfied} \\ 0.60 < x \leq 0.80 & \text{If satisfied} \\ 0.80 < x \leq 1.00 & \text{If fully satisfied} \end{cases}$$

The overall trust level $Trust_{n,t}(CSP)$ of a CSP in the TMP up to $n$ transactions in the time interval $t$ is defined as follows:

$$Trust_{n,t}(CSP) = \frac{\omega_1 \cdot SAT_{n,t}(F_{csp}) + \omega_2 \cdot Auth_{n,t}(H_{csp})}{\omega_1 + \omega_2} \quad (6).$$

Where $\omega_1$, $\omega_2$ are the weights given to the satisfaction and authentication features respectively. These parameters express the degree of influence of each feature on the overall trust value. They are defined by the $k$ CSPs participating in TMP on the interval [0, 1], according to their trust requirements

$$\omega_1 = \frac{1}{k}\sum_{i=1}^{k}\omega_{1i} \quad (7), \quad \omega_2 = \frac{1}{k}\sum_{i=1}^{k}\omega_{2i} \quad (8).$$

## V. APPLICATION EXAMPLE: IaaS CLOUD FEDERATION (CLOUD OF CLOUDS)

In this section, we present an application example in order to illustrate the feasibility of the proposed model. The particular application example focuses on the Infrastructure as a Service (IaaS) cloud federation [29], which is a partnership between different CSPs for the borrowing/lending of virtual resources including VMs, virtual clusters and virtual networks. This new perspective of cloud computing can be especially very useful to groups of small and medium cloud providers wanting to consolidate their limited resources to increase the quality of service, cost benefits, and reliability [30]. Current studies affirm that the promises of infinite computing power and infinite storage at cost which are extremely low can only be achieved by federations at the IaaS level [29], [30], [31]. Despite these promises, IaaS cloud federation is not yet mature because of many obstacles especially related to interoperability, security and trust management [30], [31]. Identity management represents the first challenge to be solved, in order to perform the authentication among heterogeneous CSPs establishing a federation [31]. The existing solutions use federated identity technologies to manage authentication and authorization across the whole cloud federation. However, as we said before, existing federation models are designed for static environments where a priori business agreements are needed among the parties and bring many security, privacy and interoperability issues. The proposed trust model is an effective method to address the aforementioned obstacles and facilitate the creation of secure IaaS cloud federations.

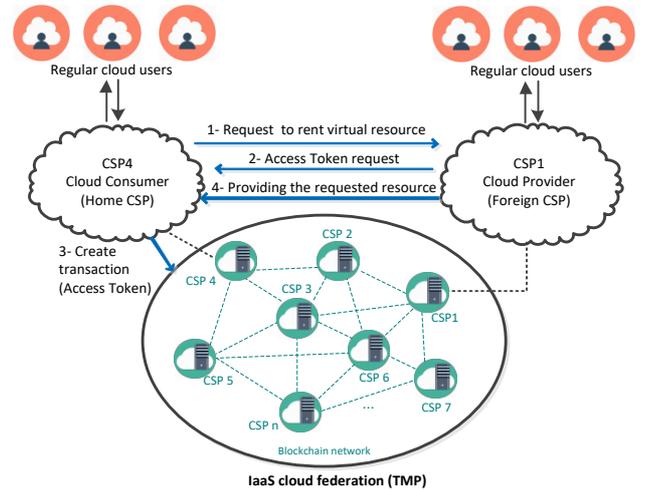

Fig. 3. Authentication scenario in order to share virtual resources.

In IaaS cloud federations, individual CSP provides and consumes virtual resources to and from other federation members. In this case, users can be either CSPs in the federation who share virtual resources from one another or regular cloud users. The home CSPs are responsible for authenticating their registered regular users and themselves when they share virtual resources from other CSPs in the federation, and foreign CSPs provide virtual resources to other CSPs in the federation or to regular users. Trust relationships between the participating CSPs in the IaaS federation are managed by the TMP, which incorporates the blockchain network and the proposed trust model. As designed, the TMP is automatically extensible, which allows other CSPs to join the IaaS federation in a dynamic way.

Figure 3 illustrates the authentication scenario when a CSP share virtual resources from another CSP in the federation. The home CSP creates an access token and sends the transaction to the blockchain network. After validating and storing the token in the blockchain, the foreign CSP provides the requested virtual resource to the home CSP.

## VI. CONCLUSIONS AND PERSPECTIVES

This paper presents a new application of the blockchain in the context of cloud identity management. The proposed model provides an authentication mechanism as well as a decentralized trust model based on the blockchain. With this trust model, CSPs do not require pre-configured trust relationships to interact, whilst trust can change dynamically according to the entities behavior. This approach presents an effective identity management solution for facilitating the creation of secure IaaS cloud federations. It also can be applied to a number of different use cases in distributed and open environments. As a future work, we expect to apply the proposed model in a real cloud environment in order to carry out tests and experiments. Moreover, we intend to conduct in-depth investigation on how to provide more desired properties, pertaining to security and usability [32]. We will rely on the Hyperledger Fabric project for implementing and testing the proposed blockchain based solution.